\documentclass[twocolumn,preprintnumbers,amsmath,amssymb]{revtex4}

\usepackage{graphicx}
\usepackage{dcolumn}
\usepackage{bm}
\usepackage{epstopdf}
\usepackage{xcolor}
\begin{document}

\title{Synergistic Laser Wakefield/Direct Laser Acceleration in the Plasma Bubble Regime}

\author{Xi Zhang, Vladimir N. Khudik, and Gennady Shvets}

\affiliation{Department of Physics and Institute for Fusion Studies, The University of Texas at Austin, Austin, Texas 78712, USA}
\date{\today}

\begin{abstract}
The concept of a hybrid laser wakefield/direct laser plasma accelerator is proposed. Relativistic electrons undergoing resonant betatron oscillations inside the plasma bubble created by a laser pulse are accelerated by gaining energy directly from the laser pulse and from its plasma wake. The resulting bifurcated phase space of self-injected plasma electrons contains a population that experiences wakefield acceleration beyond the standard one-dimensional limit because of the multi-dimensional nature of its motion that reduces the phase slippage between the electrons and the wake.
\end{abstract}

\maketitle

Advances in laser technology are transforming the idea of
laser-based acceleration of charged particles into one of the most promising high-gradient concepts~\cite{mourou_nphot13}. Broadly speaking, laser acceleration concepts can be divided into two classes: far-field
particle accelerators, where acceleration is accomplished by
transverse laser fields that do not require any external
electromagnetic structures, and near-field particle accelerators,
where the laser field is significantly modified by the presence of a linear or nonlinear medium. In a typical far-field accelerator,
such as an inverse free-electron laser\cite{babzien_prl98,steenbergen_prl96} or inverse
ion-channel laser~\cite{whittum_prl90,pukhov_dla,gahn_prl},
relativistic electrons executing undulating or betatron motion gain energy directly from the laser. On the contrary, in the near-field laser-wakefield acceleration (LWFA)~\cite{dawson_prl_lwfa} regime, the electrons gain energy indirectly from the electric field of the plasma wave which is excited by a laser pulse.

Several unique features of plasmas conspire to make the LWFA one
of the most exciting near-field acceleration concepts of the past decade~\cite{malka_nphys08,esarey_review_rmp09,hooker_nphot13}: high accelerating gradient; the available pool of electrons supplied by the plasma acting as an injector; the replaceability of the plasma accelerating structure after each laser pulse. The strongly nonlinear regime of the LWFA, corresponding to the complete blow-out of the plasma electrons from the laser's path~\cite{rosenzweig_PRA91,pukhov_bubble}, is
particularly promising for generating high-energy mono-energetic
electron beams~\cite{faure_nature04,geddes_nature04,mangles_nature04} that have recently reached GeV-scale energies~\cite{leemans,downer_nat_comm,kim_prl}.
The key enabling mechanism for narrow energy spread is the electron injection into the resulting plasma "bubble" over a short distance accomplished by engineering either the plasma density ramp
~\cite{geddes_prl08,schmid_prstab10,buck_prl13,gonsalves_nphys11,austin_pop13}
or the rapid variation of the bubble's size during
self-focusing~\cite{kalmykov_prl,downer_nat_comm} along the
laser's path. However, phase slippage (dephasing) between the electric field inside the bubble propagating with sub-relativistic speed $v_b$ and ultra-relativistic electrons co-moving with the bubble with $v_x \approx c$ limits the energy gain.

The far-field plasma-based direct laser acceleration (DLA) has also been considered in the past
~\cite{pukhov_dla,cary_prl,suk_pop,mori_ppcf,robinson_prl,arefiev_pop14},
especially in the context of developing efficient x-ray and
$\gamma$-ray radiation sources~\cite{cary_prl,phuoc_pop,phuoc_beta,jaro}. DLA occurs when the laser pulse transfers energy and momentum to relativistic electrons undergoing betatron oscillation in a
partially~\cite{gahn_prl,suk_pop,mori_ppcf} or fully~\cite{cary_prl,phuoc_pop,phuoc_beta,jaro} evacuated plasma channel. For a laser pulse with frequency $\omega_L$ and phase velocity $v_{\rm ph}$ to resonantly interact with a co-propagating electron executing betatron motion with frequency $\omega_{\beta}$, the following resonance condition must be satisfied over the length of the plasma: $\omega_d \equiv \omega_L(1-v_x/v_{\rm ph}) = \pm \omega_{\beta}$. The main limitation of the DLA is that, generally, the experimentally measured energy distribution of the accelerated electrons is Boltzmann-like~\cite{pukhov_dla,gahn_prl}. Considerable improvement in laser-plasma acceleration could be achieved if energy gains from the laser and from the wakefield were combined while maintaining (or even reducing) the narrow energy spread characteristic of self-injected bubble-regime LWFAs~\cite{downer_nat_comm}.

It is by no means obvious that such synergistic combination of the two acceleration mechanisms is possible. For example, rapid
particle acceleration by the plasma wakefield can rapidly detune
the betatron resonance, as well as damp the amplitude of the betatron motion~\cite{cary_prl} which determines DLA's accelerating gradient~\cite{pukhov_dla}. The laser pulse profile which is optimal for DLA may affect the structure of the plasma bubble, thereby reducing the energy gain from the wake and/or inhibiting self-injection. In this Letter we demonstrate that the two mechanisms can, in fact, act synergistically, with DLA significantly increasing the LWFA energy gain by extending the dephasing length. Using particle-in-cell (PIC) simulations, we predict the emergence of two distinct groups of self-injected electrons separated in time and in phase space: the high-energy DLA group that experiences large and comparable energy gain from both acceleration mechanisms, and a lower-energy non-DLA group that experiences smaller energy gain from the LWFA mechanism and no energy gain from the DLA mechanism. The larger wake acceleration experienced by the DLA population is shown to be caused by its delayed dephasing.

Before presenting the results of self-consistent PIC simulations that model all aspects of the laser evolution, electron injection, acceleration, and separation into DLA/non-DLA populations, we first develop qualitative understanding of hybrid DLA/LWFA using test-particle simulations of electron dynamics in the combined wakefield and laser fields. We adopt a simplified description~\cite{cary_prl,phuoc_pop,phuoc_beta} of the electromagnetic fields in the 2D ($x-z$) geometry. The accelerating/focusing fields of the plasma wake inside a spherical bubble with radius $r_b$ propagating with relativistic velocity $v_b \approx c(1 - 1/2\gamma_b^{2})$ are approximated as $E_{x}^{(W)} =  m\omega_p^2 (x - r_b - v_b t)/2e$, $E_{z}^{(W)} =  m\omega_p^2 z/2e$, respectively, where $\omega_p = \sqrt{4\pi e^2 n/m_e}$ is the plasma frequency, $n$ is the plasma density, and $m_e$ is the electron mass. At the bubble's center $\zeta \equiv x - v_bt =r_b$ the accelerating field changes sign.

For simplicity, the linearly polarized laser fields were assumed to be planar and given by $E_z^{(L)} = -E_0 \sin{\omega_L (t - x/v_{\rm ph})}$ and $B_y^{(L)} = B_0 \sin{\omega_L (t - x/v_{\rm ph})}$, where $B_0 = c E_0/v_{\rm ph}$. The equations of electron motion are then given by
\begin{eqnarray}
  \frac{dp_x}{dt} &=& -e \left( E_{x}^{(W)} - \frac{v_z}{c} B_y^{(L)} \right) \nonumber \\
  \frac{dp_z}{dt} &=&  -e \left( E_{z}^{(W)} + E_z^{(L)} + \frac{v_x}{c} B_y^{(L)} \right),\label{eq:model}
\end{eqnarray}
and the following laser and plasma parameters scaled to the laser wavelength $\lambda_L=2\pi c/\omega_L=0.8\mu$m were chosen for the simulations below: $\omega_p/\omega_L = 0.032$ (corresponding to plasma density $n = 1.8\times 10^{18}cm^{-3}$), $r_b = 22\lambda_L$, $\gamma_b = 18$, and $E_0 \approx 2.5 m_e c\omega_L/e$. For these parameters the peak accelerating gradient $E_{\rm max}^{(W)}$ at the back of the bubble ($x=v_b t$) is $E_{\rm max}^{(W)} \approx E_0/40 \approx 2$GV/cm. These parameters were chosen to approximately mimic the parameters of PIC simulations presented below. From Eq.~(\ref{eq:model}), the natural betatron frequency an electron with relativistic factor $\gamma$ is $\omega_{\beta} = \omega_p/\sqrt{2\gamma}$.

We first consider the case of a subluminal laser pulse with $v_{\rm ph} = 0.9985c$~\cite{cary_prl}. Although the proposed approaches to achieving $v_{\rm ph}<c$ such as using cluster plasmas~\cite{tajima_pop99}, residual non-neutral gas~\cite{hafizi_ieee00}, or corrugated plasma waveguides~\cite{york_prl08} are challenging to implement in the context of ultra-intense laser pulses, we briefly analyze the subluminal case below because it provides a stark illustration of the delayed dephasing via direct laser-electron interaction. Test electrons are injected at $t=0$ near the back of the bubble at $x = 2.65\lambda$ with a constant value of $\gamma = 25$. The initial transverse positions $z$ and momenta $p_z$ were chosen to span a wide range $0 < \epsilon_{\perp 0}/m_ec^2 < 1$ of transverse energies~\cite{kostyukov_pop04,malka_prl11} $\epsilon_{\perp} = p_z^2/2\gamma m_e + \gamma m_e \omega_{\beta}^2 z^2/2$.

\begin{figure}[ht]
\centering
   \includegraphics[height=0.3\textheight,width=0.9\columnwidth]{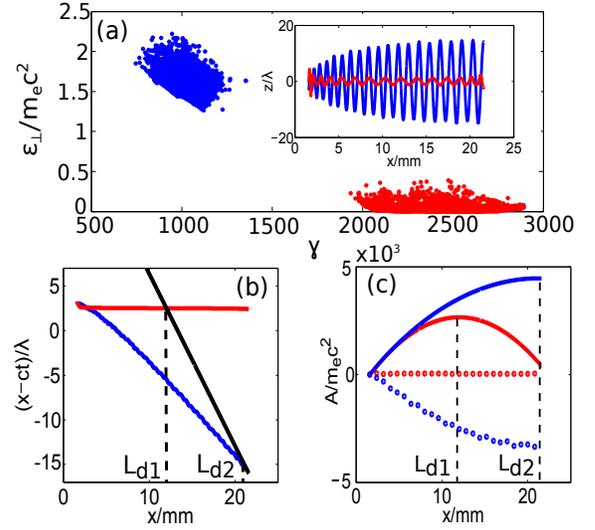}
\caption{Single-particle dynamics in combined wake/laser fields with $v_{\rm ph} < c$. (a) Fragmentation of the $(\gamma,\epsilon_{\perp})$ phase into DLD (blue) and non-DLD (red) electron populations at $x=1.3$cm. Inset: betatron trajectories of two representative electrons. (b) Longitudinal trajectories of the same electrons; black curve: bubble's center $x=v_b t$. (c) Energy gain by the two electrons from wake ($A_W$, solid lines) and from the laser ($A_L$, dashed lines).}\label{fig:sub}
\end{figure}

The bifurcated $(\gamma,\epsilon_{\perp}/m_ec^2)$ phase space of the injected test electrons after the propagation distance of $x=ct=1.3$cm is shown in Fig.~\ref{fig:sub}(a): one group of electrons (blue) gains considerable transverse energy $\epsilon_{\perp}$ from the laser while the other group (red) experiences considerable reduction in $\epsilon_{\perp}$. By following two representative electrons (one from each group, see inset), the following remarkable properties of the two groups are observed.
{\it (i) Direct Laser Deceleration (DLD)}: the work $A_L = - \int e {E}_z^{(L)} \cdot v_z dt$ done by the laser field on the first group of electrons (blue lines in Figs.~\ref{fig:sub}) is negative as shown by the dashed line in Fig.~\ref{fig:sub}(c).  The non-DLD electrons do not exchange energy with the laser pulse. The physics of the DLD is related to the anomalous Doppler effect (i.e. $- \omega_d = \omega_{\beta}$) that has been investigated in dielectric-loaded or periodically loaded waveguides~\cite{guo_prl82,jerby_pre97}. Qualitatively, if an electron interacts with the laser alone, a simple relationship between the changes in $\epsilon_{\perp}$ and $\gamma$ can be derived: $\Delta \gamma (1-c/v_{\rm ph}) = \Delta \epsilon_{\perp}/mc^2$, thus implying that DLD ($\Delta \gamma < 0$) is necessary for the resonant excitation of betatron oscillations ($\Delta \epsilon_{\perp} > 0$) whenever $v_{\rm ph} < c$. {\it (ii) Laser-delayed dephasing} is apparent  from Fig.~\ref{fig:sub}(b), where the trajectory of the DLD electron is shown to cross the bubble's center much later than that of the non-DLD electron: $L_{d2} \approx 2 L_{d1}$. The dephasing rate $d\zeta/dt$ is suppressed by the resonant excitation of the betatron oscillation according to
\begin{equation}\label{eq:dephasing}
  \frac{d\zeta}{d(ct)} \approx \frac{1}{2\gamma_b^2} - \frac{1 + \langle p_z^2/m_e^2c^2 \rangle}{2\gamma^2},
\end{equation}
where $\langle p_z^2 \rangle \approx \gamma m_e \epsilon_{\perp}$ represents the time-averaged betatron oscillation momentum. An important manifestation of the delayed dephasing for DLD electrons is that they experience much greater energy gain $A_W = - \int e {E}_x^{(W)} v_x dt$ from the wakefield (solid lines in Fig.~\ref{fig:sub}(c)) compared with non-DLD electrons. Note, however, that the total energy gain $A = A_W + A_L$ is smaller for DLD electrons because they amplify the laser pulse at the expense of the energy gained from the wake.
\begin{figure}[ht]
\centering
   \includegraphics[height=0.15\textheight,width=0.9\columnwidth]{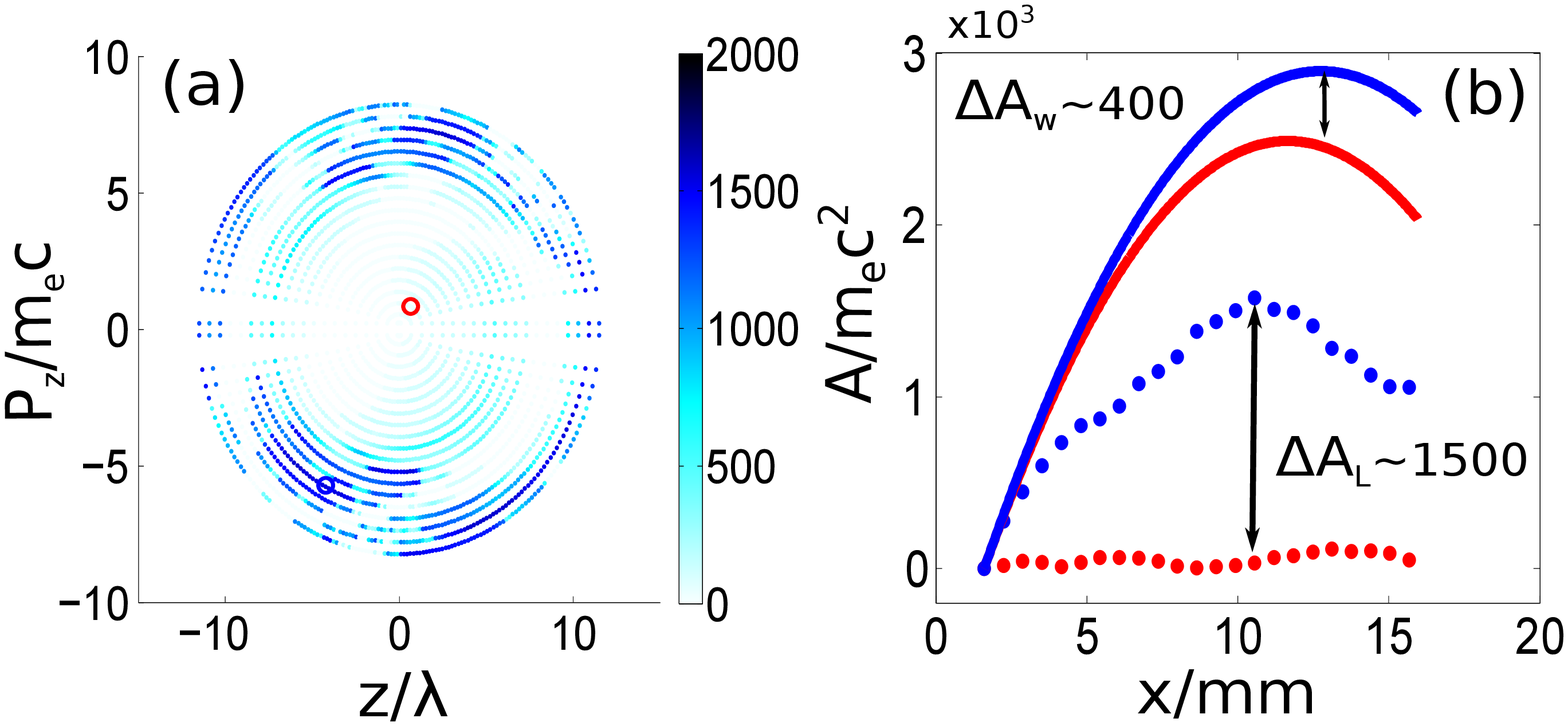}
\caption{Single-particle dynamics in combined wake/laser fields with $v_{\rm ph} > c$. (a) Color-coded laser energy gain $A_L$ as a function of the initial conditions in the $(z_0,p_{z0})$ phase space. Elliptical curves represent $\epsilon_{\perp} ={\rm const}$ curves. (b) Energy gain from the laser/wake ($A_L$: dashed lines, $A_W$: solid lines) for two test electrons with initial conditions marked in (a) by circles. Blue lines: DLA, red lines: non-DLA test electrons. All other parameters: same as Fig.~\ref{fig:sub}.}\label{fig:sup}
\end{figure}

Next, we consider a more realistic case of the superluminal phase
velocity ($v_p = 1.00036c$ corresponding to laser propagation in plasma with $n = 1.8\times 10^{18}$cm$^{-3}$; all other laser/wake parameters and initial conditions of the test electrons are the same as in the subluminal case). In the $v_{\rm ph} > c$ case the electrons gaining transverse energy are also gaining energy from the laser, i.e. $A_L > 0$. It is apparent from Fig.~\ref{fig:sup}(a) that, while $A_L$ depends on the initial
phase of the electron's betatron oscillation (i.e. on the specific values of  $p_{z0}$ and $z_0$), a large initial value of the transverse energy is a pre-condition for DLA.

Laser and wake energy gains of two representative DLA (blue) and non-DLA (red) electrons with initial transverse energies $\epsilon_{\perp 0} = 0.8 m_ec^2$ and $\epsilon_{\perp 0} = 0.1 m_ec^2$), respectively, are compared in Fig.~\ref{fig:sup}(b).
The synergistic nature of the hybrid DLA/LWFA is apparent: the DLA electron gains more energy from the wake than a non-DLA electron, with the difference of $\Delta A_W \approx 0.2$GeV) being due to delayed dephasing. At the same time, the DLA electron gains $A_L \approx 0.7$GeVs energy from the laser, thereby almost doubling its total final energy $\epsilon_{\rm tot} \equiv \gamma m_ec^2$ compared with its non-DLA counterpart.

\begin{figure}[ht]
\begin{center}
\vspace{2 mm}
\includegraphics[height=0.3\textheight,width=0.9\columnwidth]{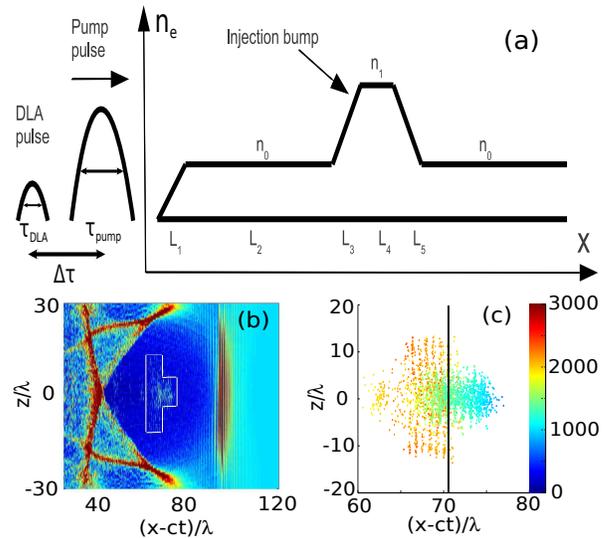}
\end{center}
\caption{(a)Schematic representation of the laser pulse format and plasma density profile. (b) Plasma electron density in the bubble regime at $x = 1$cm; self-injected electron bunch inside the white box has advanced approximately to the middle of the bubble. (c) Zoom-in of the self-injected electrons color-coded according to their relativistic factor $\gamma$; black vertical line: bubble's center. Plasma parameters: $L_1 = L_3 = L_4 = L_5 = 0.1$mm, $L_2 = 1.6$mm; $n_0 = 1.8\times 10^{18}cm^{-3}$, $n_1 = 3n_0$. Laser parameters: $I_{\rm pump} = 2\times 10^{19}$W/cm$^2$, $I_{\rm DLA} = I_{\rm pump}/5$, pulse durations $\tau_{\rm pump} =50$fs and $\tau_{\rm DLA} = 30$fs, inter-pulse time delay $\Delta \tau = 67$fs.}\label{fig:densprof}
\end{figure}

Based on the results of single-particle modeling, we can now
formulate the conditions for achieving synergistic DLA/LWFA in a
realistic laser-plasma accelerator. {\it First,} considerable overlap between laser field and injected electrons is required for effective DLA. {\it Second,} electrons must be injected into the bubble with large transverse energy. We use a 2D PIC code VLPL~\cite{pukhov_vlpl} to model the self-consistent interaction
of a multi-terawatt laser pulse with tenuous ($n = 1.8\times
10^{18}cm^{-3}$) plasma to demonstrate that these two conditions
can be met.  The first condition is satisfied by employing two laser pulses (labeled as pump and DLA in Fig.~\ref{fig:densprof}(a); see caption for laser/plasma
parameters), where a much weaker time-delayed DLA pulse has no
observable effect on the bubble shape and accelerating field, yet
enables DLA by overlapping with self-injected electrons.

The second condition is met by engineering the self-injection of
the background plasma electrons into the bubble. A short injection density bump shown in Fig.~\ref{fig:densprof}(a) is utilized to
rapidly deform the plasma bubble, thereby causing
self-injection~\cite{kalmykov_prl,austin_ppcf,austin_pop13,pak,suk_prl,malka_prl13}
of plasma electrons. Note that, although the bubble is fully
formed for $x<L_1+L_2$, no self-injection occurs prior or after
the laser encountering the density bump. Experimental approaches
to generating such density bumps have been described
elsewhere~\cite{pai05,pai06}. The bump-facilitated injection can
be thought of as a less "gentle" version of transverse
injection~\cite{malka_prl13} that imparts self-injected
electrons with large transverse energy $\epsilon_{\perp}$ needed for efficient DLA as illustrated in Fig.~\ref{fig:sup}(a).

As the injected electrons, shown in Fig.~\ref{fig:densprof}(b)
after propagating for $x=1$cm through the plasma, advance towards
the center of the bubble and experience dephasing, a clear
separation into DLA and non-DLA groups occurs. Electrons color-coded according to their final energy are shown in
Fig.~\ref{fig:densprof}(c), which is a zoom-in of
Fig.~\ref{fig:densprof}(b) in the vicinity of the bubble's center
indicated by a vertical black line. Clearly, the highest energy
electrons comprising the DLA group have a much larger betatron
oscillation amplitude, and are spatially located behind the
lower-energy non-DLA group of electrons. According to
Eq.~(\ref{eq:dephasing}), DLA electrons advance slower through the bubble because they have much higher transverse momentum (up to
$p_z=100m_ec$) imparted directly by the DLA pulse.
\begin{figure}[ht]
\centering
   \includegraphics[height=0.3\textheight,width=0.9\columnwidth]{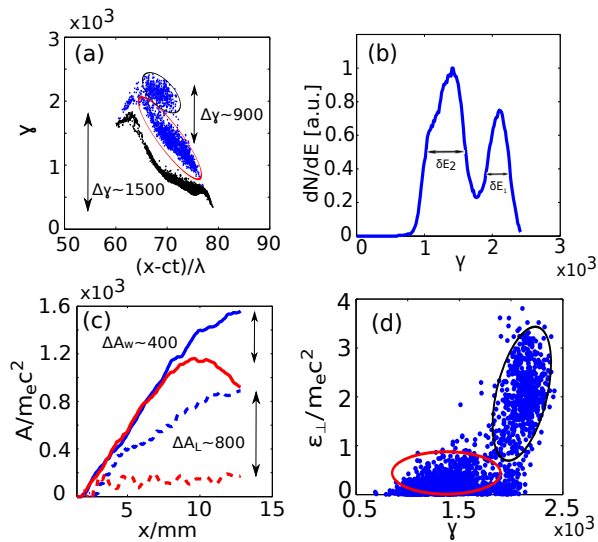}
\caption{(a) Phase space of self-injected electrons for double-pulse (blue dots) and single-pulse (black dots) laser formats. (b) Energy spectrum for double-pulse (pump + DLA) formats. Energy spreads: $\delta E_1 \simeq 350 m_ec^2$, $\delta E_2 \simeq 600 m_ec^2$. (c) Energy gain from the wake ($A_W$: solid lines) and laser ($A_L$: dashed line) fields for DLA (blue) and non-DLA (red) representative electrons. (d) Bifurcated phase space $(\gamma,\epsilon_{\perp})$ shows correlation between total and transverse energies for DLA electrons.}\label{fig:density}
\end{figure}

The bifurcated $(x-ct,\gamma)$ phase space and the total energy
spectrum of the accelerated electrons are plotted in Figs.~\ref{fig:density}(a,b), respectively (blue-colored). The DLA (black-circled) and non-DLA (red-circled) electrons are clearly
separated in energy and space, with their energy spectra peaking
at $\epsilon^{\rm DLA}_{\rm tot} = 1.1$GeV and $\epsilon^{\rm
n-DLA}_{\rm tot} = 0.65$GeV, respectively. To illustrate the role
of the time-delayed DLA laser pulse on phase space bifurcation, we carried out PIC simulations for the single-pulse LWFA case, i.e.
with the same bubble-producing pump pulse ($I_{\rm pump} = 2\times 10^{19}$W/cm$^2$ corresponding to $a_{\rm pump} = 3$) but no DLA
pulse. The resulting electron phase space shown in Fig.~\ref{fig:density}(a) (black dots) do not show any phase space fragmentation, thus indicating that no DLA electrons are produced.

The synergistic nature of the DLA/LWFA mechanisms can be
demonstrated by comparing the LWFA gains $A_W$ plotted in
Fig.~\ref{fig:density}(c) for two representative DLA and non-DLA
electrons. The non-DLA electron gains less energy, and promptly moves into the decelerating phase of the bubble's field
(red solid line), while the DLA electron gains more energy and
does not experience dephasing (blue solid line). At the same time, the DLA electron gains considerable energy ($A_L \approx
900m_ec^2$) directly from the laser. The combination of larger
gains from the wake ($\Delta A_W \approx 400m_ec^2$) and from
the laser ($\Delta A_L \approx 800 m_ec^2$) explains why DLA
electrons acquire much higher energy than non-DLA electrons.

Another intriguing difference between the two groups of electrons
is observed by plotting the $(\gamma,\epsilon_{\perp})$ phase
space in Fig.~\ref{fig:density}(d). While there is no correlation
between $\gamma$ and $\epsilon_{\perp}$ for the non-DLA group, a
strong positive correlation exists for the DLA group. A
relativistic beam with finite emittance and energy spread
possessing such correlation between total energy and transverse
action of its electrons is referred to as
conditioned~\cite{sessler}. It has been suggested that beam
conditioning \cite{sessler,sprangle,schroeder,zholents,penn} can
considerably improve gain and efficiency of FELs if the
correlation between $\epsilon_{\perp}$ and $\gamma$ is such that
any deviation of individual electron's energy $\Delta \gamma_i =
\gamma_i - \gamma_d$ from the design energy $\gamma_d$ is
compensated by the corresponding increase in its transverse energy $\epsilon_{\perp i}$, so that there is no spread in the
longitudinal velocity $\Delta v_{x i}$ is minimized.

In conclusion, we have proposed and theoretically demonstrated a new type of a plasma-based accelerator: a hybrid laser wakefield/direct laser accelerator. The synergistic nature of the LWFA/DLA mechanism manifests itself in compounding the distinct energy gains from the plasma wake and directly from the laser pulse while increasing the former because of the delayed dephasing caused by the latter. Phase space bifurcation of the self-injected electrons into two distinct groups of high-energy DLA and lower-energy non-DLA particles is demonstrated. Future work will explore the possibility of developing incoherent and coherent (e.g., FELs) radiation sources based on DLA electrons.

This work was supported by DOE grants DE-SC0007889 and DE-SC0010622, and by an AFOSR grant FA9550-14-1-0045.

\end{document}